\documentclass[pra,twocolumn,floatfix,superscriptaddress,longbibliography,notitlepage]{revtex4-2}

\usepackage[utf8]{inputenc}
\usepackage{mathtools}
\usepackage{amsfonts}
\usepackage{amssymb}
\usepackage[colorlinks]{hyperref}
\usepackage[noabbrev]{cleveref}
\usepackage{listings, color}
\usepackage{enumitem}
\usepackage{float}
\usepackage{textcomp}
\usepackage{graphicx}
\usepackage[table,xcdraw]{xcolor}
\usepackage{appendix}
\usepackage{todonotes}
\usepackage{ragged2e}
\usepackage{placeins}
\graphicspath{{Figures/}}
\newcommand{\COMMENT}[1]{}

\begin{document}
\title{Probabilistic Flux Limiters}

\author{Nga Nguyen-Fotiadis}
\affiliation{Information Sciences, CCS-3, Los Alamos National Laboratory, Los Alamos, NM 87545, USA}

\author{Robert Chiodi}
\affiliation{Computational Physics and Methods, CCS-2, Los Alamos National Laboratory, Los Alamos, NM 87545, USA}

\author{Michael McKerns}
\affiliation{Information Sciences, CCS-3, Los Alamos National Laboratory, Los Alamos, NM 87545, USA}

\author{Daniel Livescu}
\affiliation{Computational Physics and Methods, CCS-2, Los Alamos National Laboratory, Los Alamos, NM 87545, USA}

\author{Andrew Sornborger}
\email{sornborg@lanl.gov}
\affiliation{Information Sciences, CCS-3, Los Alamos National Laboratory, Los Alamos, NM 87545, USA}

\begin{abstract}
  \noindent
  The stable numerical integration of shocks in compressible flow simulations relies on the reduction or elimination of Gibbs phenomena (unstable, spurious oscillations). A popular method to virtually eliminate Gibbs oscillations caused by  numerical discretization in under-resolved simulations is to use a flux limiter.  A wide range of flux limiters has been studied in the literature, with recent interest in their optimization via machine learning methods trained on high-resolution datasets. The  
  common use of flux limiters in numerical codes as plug-and-play blackbox components makes them key targets for design improvement. Moreover, while aleatoric (inherent randomness) and epistemic (lack of knowledge) uncertainty is commonplace in fluid dynamical systems, these effects are generally ignored in the design of flux limiters. Even for deterministic dynamical models, numerical uncertainty is introduced via coarse-graining required by insufficient computational power to solve all scales of motion.
Here, we introduce a conceptually distinct type of flux limiter that is designed to handle the effects of randomness in the model and uncertainty in model parameters.  This new, {\it probabilistic flux limiter}, learned with high-resolution data, consists of a set of flux limiting functions with associated probabilities, which define the frequencies of selection for their use. Using the example of Burgers' equation, we show that a machine learned, probabilistic flux limiter may be used in a shock capturing code to more accurately capture shock profiles.  In particular, we show that our probabilistic flux limiter outperforms standard limiters, and can be successively improved upon (up to a point) by expanding the set of probabilistically chosen flux limiting functions.
\end{abstract} 
\maketitle

\section{Introduction}
Numerical methods for simulating fluid flows are inherently limited by the need to discretize the originally continuous flow equations. Since fully resolving all the dynamically relevant spatio-temporal scales is not feasible for most practical applications on today's computers, fluid dynamics computations are generally limited to the use of coarse meshes.  

One of the most evident drawbacks of using coarse meshes arises when shocks form in compressible flows. The physical width of the shock may be orders of magnitude smaller than the mesh spacing, which results in spurious oscillations in solution fluxes about the shock, unless an additional dissipation mechanism is provided to artificially widen the shock so it becomes representable on the actual mesh. This problem becomes more acute for higher order schemes, which are desirable in smoother regions of the flow due to their lower truncation errors, but which have less dissipation to regularize the coarse mesh equations around the shock. This type of oscillation due to the Gibbs effect \cite{wilbraham1848certain}, does not occur for first-order flux approximations.


%

In order to resolve the tension between the benefit of using high-order derivatives when possible with the need to reduce the Gibbs effect, flux limiters were introduced in so-called shock capturing methods \cite{godunov1959difference,van1979towards,colella1984piecewise,harten1997high,shu1988efficient}. A flux limiter, $\phi$, interpolates between low- and high-order derivatives depending on the flux ratio between grid points in a numerical flow simulation, $r_i = \frac{u_i - u_{i-1}}{u_{i+1} - u_i}$, (here, $u_i$ denotes a flow variable at grid location $i$). In regions of sharp differences between flow variables, first-order differences are favored in the interpolation to reduce spurious oscillations, but in smooth solution regions, high-order derivatives are favored to provide more accurate computation of flow quantities.

\begin{figure}[h]%
    \includegraphics[width=0.4\textwidth]{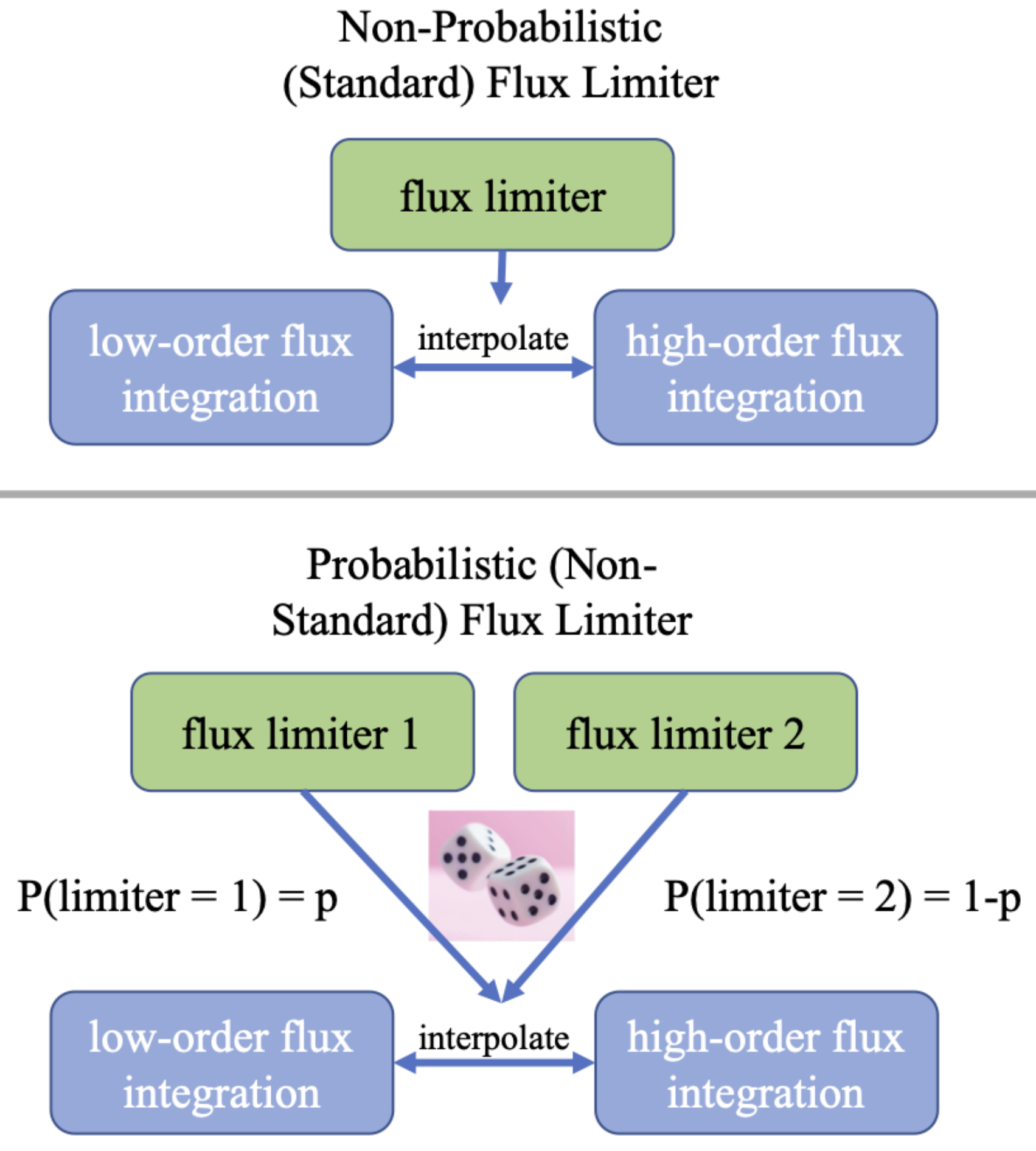}
    \caption[justification=left]{A probabilistic flux limiter expands the flux limiter concept from an individual, deterministic interpolating function (top panel) to a set of interpolating functions applied probabilistically with probabilities drawn with replacement from a distribution learned from high-resolution data (bottom panel). In the example shown here, the probabilistic limiter consists of two flux limiters randomly selected with probability $p$ and $1-p$ for each flux computation.}
    \label{fig:ProbFL}
\end{figure}

A wide range of flux limiters has been studied in the literature \cite{zhang2015review}. Many of these limiters were designed to fit within the 2nd-order TVD region \cite{sweby1984high}. Full containment of a flux limiter within the 2nd-order TVD region is a sufficient, but not necessary, condition to eliminate the possibility of Gibbs effects in 2nd-order shock capturing schemes (see discussion around Eqs.~2.15 and 2.16 in \cite{sweby1984high}). Some standard flux limiters, such as the van Albada limiter, work well, but do not fit within this region. Similarly, machine learned flux limiters for the coarse-grained Burgers' equation have a unique functional appearance and lie outside the 2nd-order TVD region, yet still outperform other limiters in accuracy \cite{NguyenMcKernsSornborger}. All limiters previously considered in the literature \cite{zhang2015review,NguyenMcKernsSornborger} consist of a single flux limiting function, $\phi(r)$, used to deterministically interpolate between first- and high-order fluxes.

\begin{figure}[h]%
    \includegraphics[width=0.5\textwidth]{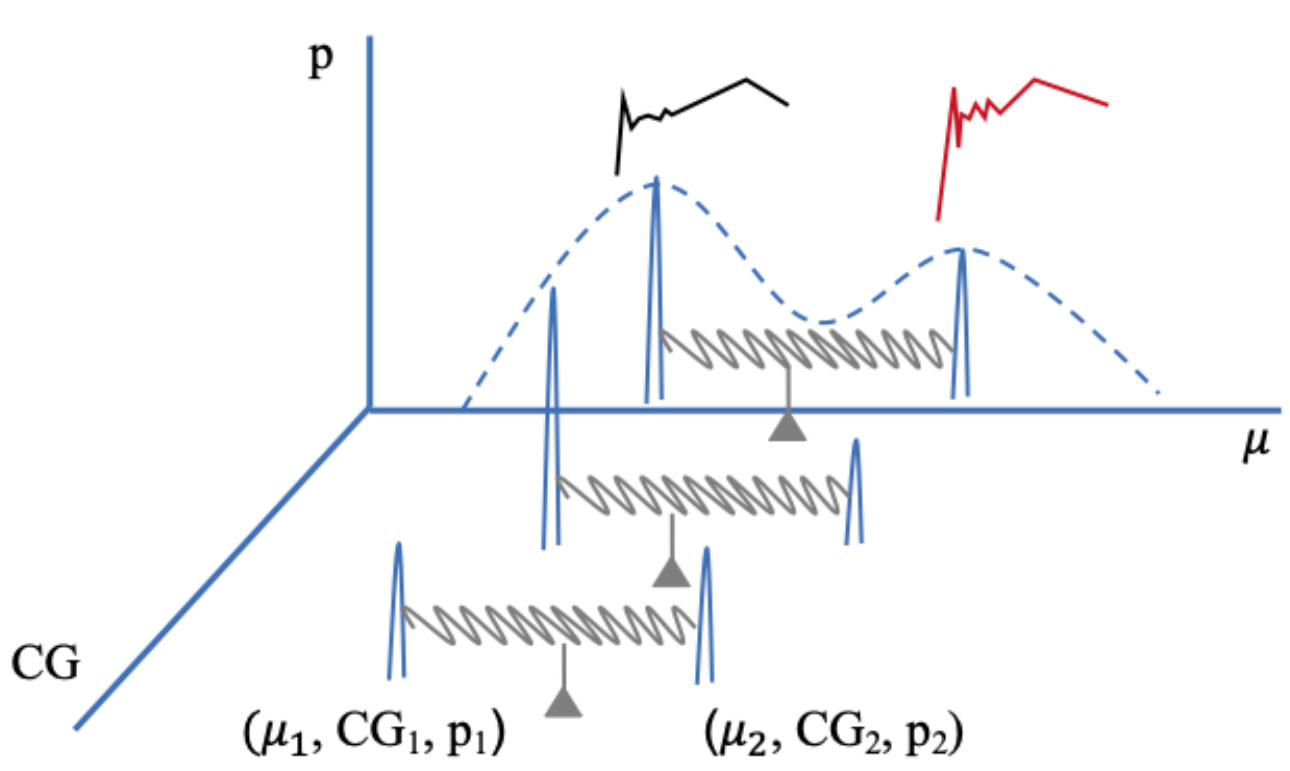}
    \caption[justification=left]{Cartoon of the hypercube of inputs for the optimization of a probabilistic flux limiter. A mean constraint on viscosity ($\mu$) is visualized as a fulcrum, and a variance constraint on $\mu$ is seen as a spring. Inputs to a flux limiter are specified by coarse graining ($CG$), total degrees of freedom of the limiter ($K$), and probability-weighted viscosity $(\mu, p)$. In this figure, the probability distribution of possible flux limiters is modeled by a discrete distribution of $\mu$ composed of $N_D = 2$ Dirac delta functions optimized over the parameter space $(CG, K, \mu, p)$ with $\sum_{m}^{N_D} p_m = 1$. In the text, we explore distributions of $\mu$ with up to $N_D=3$.}
    \label{fig:ProbFLUQ}
\end{figure}

Our primary contribution in this article is the introduction of a new, probabilistic conception of flux limiters. 

In the presence of incomplete, imperfect knowledge (sometimes called epistemic uncertainty) about the 
operators 
involved in integrating non-linear flows, and especially in a high-consequence decision-making context, it makes sense to adopt a conservative posture to uncertainty quantification (UQ).
This, in turn, naturally leads to UQ being posed as an optimization or machine learning problem, as this can facilitate estimates of bounds on the quantities of interest in the system, in our case a flux limiter.

Often, the UQ objective is to determine or estimate the expected value of some measurable quantity of interest, given an input distribution and a response function. However, in practice, the true response function and input distribution are rarely known precisely. 
Commonly, there may be some knowledge about the probability distribution and response function (perhaps through measurements performed with some degree of statistical confidence) such that the true input distribution and response function are bounded by knowledge of the system.
This information can then be encapsulated in a probability measure — what a Bayesian probabilist would call a prior — so that we can perform (an optimization or) sampling to (calculate or) estimate (bounds on) the expected value of the quantity of interest by transforming the problem coordinates into a hypercube that includes the original coordinates $\mathcal{X}$ and the probability $\mathcal{P}$ associated with the position on $\mathcal{X}$ as defined by the probability measure \cite{OSSMO:2011, MOSSO:2010}.

With this in mind, here we consider coarse-grained flow simulations to be non-deterministic, since they necessarily ignore subgrid information. Below, we discuss a framework for implementing a type of shock capturing method that uses a (simple) probabilistic structure for a shock capturing scheme, where a fluid simulation is thought of as a probabilistic operator on data sampled from a distribution.

Specifically, we introduce a new type of flux limiter -- a probabilistic flux limiter (see Fig.~\ref{fig:ProbFL}). 
We have taken the approach of using concepts from uncertainty quantification to learn optimal flux limiters in a Monte Carlo context. The resulting probabilistic flux limiter consists of a set of flux limiters with associated probabilities, $\{ (\phi_m, p_m) : m \in [1, \dots, N_D] \}$, where $\sum_{m}^{N_D} p_m = 1$. In contrast to standard, deterministic flux limiters (Fig.~\ref{fig:ProbFL}, top panel), each evaluation of a probabilistic flux limiter (Fig.~\ref{fig:ProbFL}, bottom panel) randomly selects one of $N_D$ limiters from a probability distribution to be used in the flux computation. 

Below, with the example of Burgers' equation \cite{burgers1948mathematical,eberhard1942partial,cole1951quasi}, we demonstrate that probabilistic flux limiters may be learned for coarse-grained fluid simulations from high-resolution data. We further show that they outperform other non-probabilistic flux limiters from the literature, including deterministic machine learned flux limiters.

\section{Methods} 
A detailed framework for our machine learned flux limiter theory was introduced in~\cite{NguyenMcKernsSornborger}. Below, we summarize the deterministic, second-order shock capturing method that we used, show how we parameterized the flux limiter, and how we optimized the discretized limiter. We then go on to show how to modify this approach for a probabilistic flux limiter.

For low- and high-resolution, respectively, we choose Lax–Friedrichs (LF)
\begin{eqnarray}
    f^\mathrm{low}_{i \pm \frac{1}{2}} = f^{\mathrm{LF}}_{i \pm \frac{1}{2}} =&& \frac{1}{2} [ F(u_i) + F(u_{i+1}) \mp \alpha \frac{\Delta x}{\Delta t}(u_{i\pm 1} -u_i) ]; \nonumber \\ &&\alpha  = \max_{{u}} \left|\frac{\partial F}{\partial {u}}\right|  
    \label{eq:Lax-F}
\end{eqnarray}
and Lax-Wendroff (LW) fluxes
\begin{eqnarray}
    f^\mathrm{high}_{i \pm \frac{1}{2}} = f^{\mathrm{LW}}_{i \pm \frac{1}{2}} &&= \frac{1}{2} [F(u_{i}) + F(u_{i+1}) \nonumber \\ 
            &&\mp \frac{\Delta t}{\Delta x} \left(\frac{\partial F}{\partial u} (u_{i \pm \frac{1}{2}})\right) ( F(u_{i \pm 1}) - F(u_{i}))  ],
    \label{eq:Lax-W}
\end{eqnarray} 
where $F = \frac{u^2}{2} - \nu \frac{\partial{u}}{\partial{x}}$ is the flux defined for Burgers' equation.  We now write the conservative form of Burgers' equation as:
\begin{equation}
    u_i(t_{n+1}) = u_i(t_n)  - \frac{\Delta{t}}{\Delta x} {\Delta{F}}^i
\end{equation}
with
\begin{eqnarray}
    \label{eq:DeltaF}
    {\Delta F}^i &=& {\Delta{F}_1}^i + \phi(r_i){\Delta{F}_2}^i + \phi(r_{i-1}){\Delta{F}_3}^i \; ,
\end{eqnarray}
where
${\Delta{F}_1}^i$, ${\Delta{F}_2}^i$, and ${\Delta{F}_3}^i$ are written explicitly as:
\begin{eqnarray}
    {\Delta{F}_1}^i &=& f^{\mathrm{LF}}_{i+\frac{1}{2}} - f^{\mathrm{LF}}_{i-\frac{1}{2}}, \nonumber \\ 
    {\Delta{F}_2}^i &=&f^{\mathrm{LW}}_{i+\frac{1}{2}} - f^{\mathrm{LF}}_{i+\frac{1}{2}},  \nonumber \\
    {\Delta{F}_3}^i &=& -(f^{\mathrm{LW}}_{i-\frac{1}{2}} - f^{\mathrm{LF}}_{i-\frac{1}{2}}). 
    \label{eq:3DeltaF}
\end{eqnarray}

We discretize the flux-limiter that we will optimize, $\phi (r)$, in piecewise linear segments, where the $k$'th segment has the form, 
\begin{eqnarray}
    \phi_k (r)  &&= \phi_0 + b_1 (r_2 - r_1) + b_2 (r_3 - r_2) + ... + b_k (r - r_k) \nonumber \\ 
                &&+ 0_{k+1} + ... +0_{K},
    \label{eq:pw-linear}
\end{eqnarray}
and $r \in [r_k, r_{k+1})$, $k \in \{1, \dots, K\}$, $\phi_0 = 0$, and $b_i$ are slope coefficients. Note that for $r \le 0$, $\phi(r) = 0$ and for $r = r_K$, all terms in Eq.~\eqref{eq:pw-linear} are non-zero. Below, we use vector notation, $\boldsymbol{b} = [b_1, b_2, ..., b_k, b_{k+1}, ..., b_K]^{\mathrm{T}}$ for slope coefficients.  Eq.~\eqref{eq:pw-linear} can be rewritten as $ \phi_k(r) = \boldsymbol{b}^{\mathrm{T}} \boldsymbol{\Delta{r}}_k $
with $\boldsymbol{\Delta{r}}_k$ defined as
\begin{equation}
\boldsymbol{\Delta{r}}_k = [r_2 - r_1, r_3 - r_2, ..., r - r_k, 0,..., 0]^{\mathrm{T}}.
\label{eq:Deltar}
\end{equation}

To optimize the discretized flux-limiter in Eq.~\eqref{eq:pw-linear}, we define the mean squared error between $N$ input-output pairs, $\{ o_i(\{ u_{c}^i \}), g_i \}$:
\begin{equation}
    C = \frac{1}{2} \sum_{i=1}^{N} { ( o_i(\{ u_{c}^i \}) - g_i  )^2 }
    \label{eq:loss}
\end{equation}
as the cost. Here, $g_i$ is the high-resolution fluid velocity at the $i$-th grid position at time $t_{n+1}$ and $o_i$ is the shock-capturing method's prediction of the fluid velocity at time $t_{n+1}$ from data at the previous timestep. $o_i$ is a functional of a subset of data points $\{ u_c^i \} = \{u_{c}^{i1}, u_{c}^{i2}, u_{c}^{i3}, ..., u_{c}^{iN_c} \}$ indicated relative to the $i$-th grid position at time step $t_n$.  Here, we used $N_c = 6$ data points at time $t_n$ to predict a data point $g_i$ at $t_{n+1}$, i.e. $ \{ u_c^i \} = \{ u_{i-3}, u_{i-2}, u_{i-1}, u_{i}, u_{i+1}, u_{i+2} \}$.  Thus, $o_i(\{ u_c^i\})$ is the integration obtained with the flux-limiter method defined in Eqs.~(\ref{eq:Lax-F}), ~(\ref{eq:Lax-W}), ~(\ref{eq:pw-linear}) given a set of 6-points $\{u_c^i \}$:
\begin{equation}
    o_i (\{ u_c^i\}, t_{n+1}) = u_i(t_n) - \frac{\Delta t}{\Delta x} \Delta F^i (\{ u_c^i \}, \{b_i \}, t_n).
    \label{eq:o_i}
\end{equation}
Here, $\Delta F^i (\{ u_c^i \}, \{b_i \}, t_n)$, defined via Eqs.~\eqref{eq:DeltaF} and \eqref{eq:3DeltaF}, 
is the difference of the two fluxes.  The minimum of the cost function, Eq.~\eqref{eq:loss}, can be computed exactly by finding the unique root, $\boldsymbol{b}$, of the equation $ \frac{\partial L}{\partial {\boldsymbol{b}}} = \boldsymbol{0} $, that is:
\begin{equation}
    \sum_{i=1}^{N} \left(u_i - g_i - \frac{\Delta{t}}{\Delta{x}} \Delta{F}^i\right) \left(-\frac{\Delta{t}}{\Delta{x}}\right) \boldsymbol{\Delta{s}}_i \boldsymbol{\Delta{F}}^{i}_{2,3} = \boldsymbol{0}.
    \label{eq:root}
\end{equation}
In Eq.~\eqref{eq:root}, $\Delta{F}^i = \Delta F (\{ u_c^i \}, \{b_i \}, t_n)$ is defined via Eqs.~\eqref{eq:DeltaF} and ~\eqref{eq:3DeltaF}.  $\boldsymbol{\Delta{s}}_i = [\boldsymbol{\Delta{r}}_i, \boldsymbol{\Delta{r}}_{i-1} ]$ is a $K \times 2$ matrix with $\boldsymbol{\Delta{r}}_i$ defined in Eq.~\eqref{eq:Deltar}. $\boldsymbol{\Delta{F}}_{2,3}^i = [ \Delta{F}^i_2, \Delta{F}_3^i ]^{\mathrm{T}}$ with components $\Delta{F}_{2}^i$ and $\Delta{F}_{3}^i$ defined via Eq.~\eqref{eq:3DeltaF}.  Solving Eq.~\eqref{eq:root} reduces to solving a linear equation $ \boldsymbol{A} \cdot \boldsymbol{b} = \boldsymbol{C} $ that yields $\boldsymbol{b} = \boldsymbol{A}^{-1} \cdot \boldsymbol{C}$. Here, $\boldsymbol{A} = \boldsymbol{\Delta{r_F}} \cdot (\boldsymbol{\Delta{r_F}})^{\mathrm{T}} $ and $\boldsymbol{C} = \frac{\Delta{x}}{\Delta{t}}\sum_{i=1}^{N} { {{O_G}^i} \boldsymbol{\Delta{r_F}}^i }$, where  $\boldsymbol{\Delta{r_F}}$ is a $K \times N$ matrix with each column $\boldsymbol{\Delta{r_F}}^i$ a $K \times 1$ vector defined as $\boldsymbol{\Delta{r_F}}^i = (\boldsymbol{\Delta{s}}_i) (\boldsymbol{\Delta{F}}_{2,3}^i)$. Finally, ${O_G}^i = u_i - g_i - \frac{\Delta{t}}{\Delta{x}} \Delta{F}_1^i$.  Note that $\Delta{F}_1^i$ is defined via Eq.~\eqref{eq:3DeltaF} and we recall that $K$ is the size of the discretized flux limiter (i.e. the size of $\boldsymbol {b}$).  Hence, each matrix $\boldsymbol A$ (or $\boldsymbol C$) is a function of $N$ training data points.  We chose to discretize the flux limiter such that each segment contained an equal number of training data points.  

\begin{figure*}[ht!]%
    \includegraphics[width=0.8\textwidth]{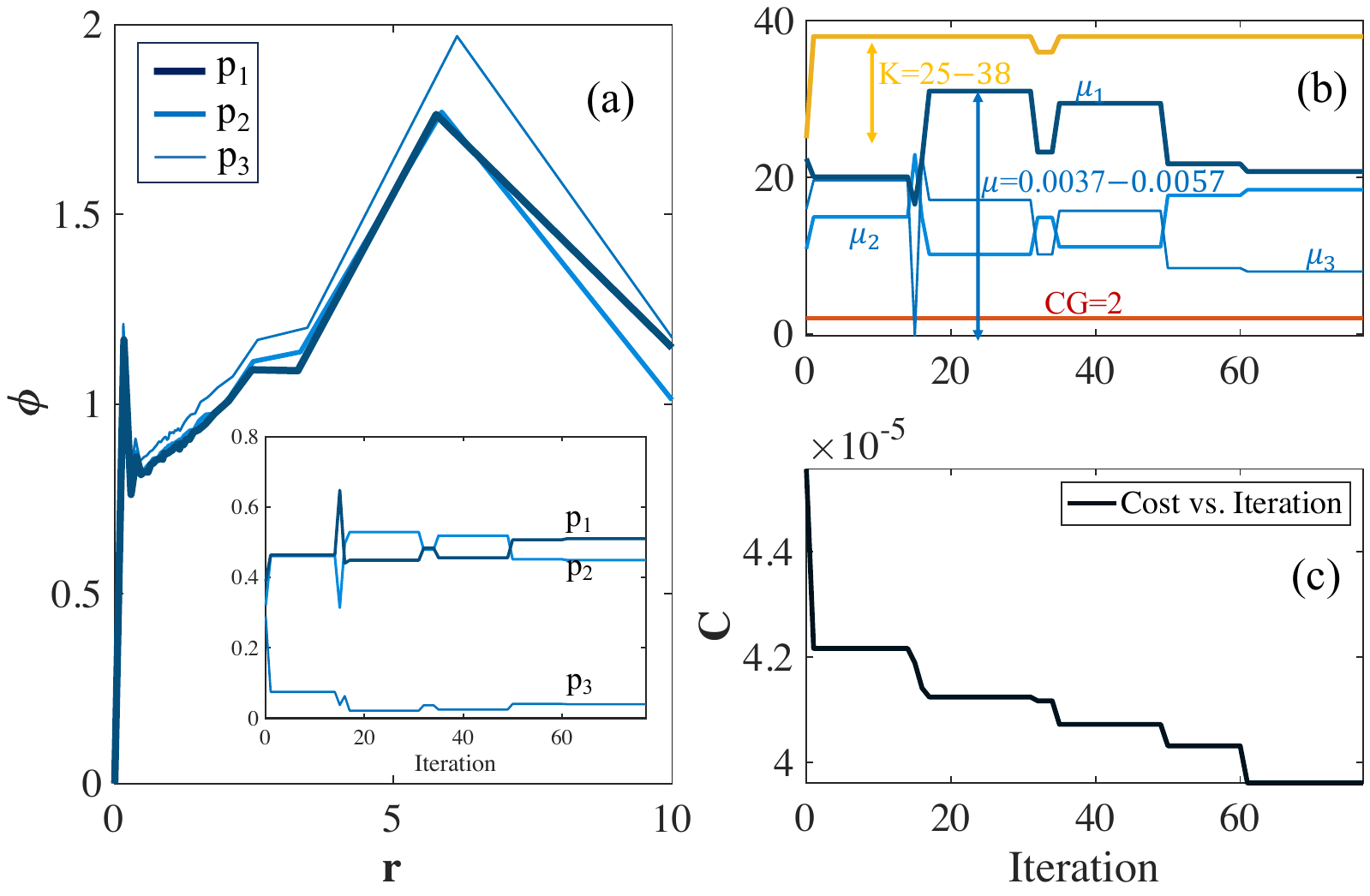}
    \caption[justification=left]{Optimization of a probabilistic flux limiter with $N_D = 3$.  (a) The optimized forms of the $3$ flux limiting functions $\phi_1$, $\phi_2$, and $\phi_3$ after $60$ iterations, the thicker the line the larger the probability. The probability distribution of the $3$ corresponding viscosity values, $\mu_1, \mu_2$, and $\mu_3$ (also shown in (b)) as a function of training iteration are depicted in the inset. (b) A plot of the $3$ viscosity values $\mu_1, \mu_2$, and $\mu_3$, bin number $K$, and coarse graining $CG$ versus iteration step. (c) The value of the cost function versus iteration step.   Optimization was performed on the entire parameter space $CG=[1,2,\dots,10], K=[2,3,\dots,38], \mu=[0.002, 0.03$] with $p=p_1 + p_2 + p_3 = 1$.}
    \label{fig:threediracsrun}
\end{figure*}
In \cite{NguyenMcKernsSornborger}, we investigated the machine learning of deterministic flux limiters and found solutions obtained by optimizing with respect to a range of parameters, including coarse graining, the total degrees of freedom of the limiter, and the viscosity, $(CG, K, \mu)$. This was essentially a meta-analysis given the costs, $C$ (\ref{eq:loss}), across the full parameter ranges.
Below, we extend this machine learning approach to optimizing probabilistic flux limiters (see Fig.~\ref{fig:ProbFLUQ}).

If we define our coordinate space, $\mathcal{X}$, by $(CG, K, \mu)$, then transformation into measure space gives us $(\mathcal{X}, \mathcal{P})$, with $\mathcal{P}$ defined by $(P_{CG}, P_K, P_{\mu})$.
In this approach, our previous, deterministic flux limiters \cite{NguyenMcKernsSornborger} were described by a probability distribution that was composed of a single Dirac delta function in each direction (and hence was deterministic). The number of Dirac delta functions used was $N_D = (1, 1, 1)$, or more compactly, $N_D = 1$.

In the current work, we extend our flux limiters to have a probabilistic nature by defining our probability measure to be composed of up to $N_D = 3$ Dirac delta functions per direction. As we will assume we have incomplete information on $\mu$, but can specify $CG$ and $K$ exactly, we have $N_D = (1, 1, 3)$. From here onward, as we only have uncertainty in $\mu$, we will use the compact form of $N_D$, and will denote $\mathcal{P}$ simply with $p \coloneqq P_{\mu}$ as a fourth coordinate.
The resulting probabilistic flux limiter can then be thought of as a set of piecewise-linear flux limiters with associated selection probabilities, $\{ (\phi_m, p_m) : m \in [1, \dots, N_D] \}$, with $\sum_{m}^{N_D} p_m = 1$. Each limiter $\phi_m \coloneqq \phi_m(CG, K, \mu)$ can then be optimized on the coordinate hypercube defined by $(CG, K, \mu, p)$. 

For the probabilistic flux limiters, $\{ (\phi_m, p_m) : m \in [1, \dots, N_D] \}$, that we consider here, with $N_D = 1,2,3$, we have
\begin{equation}
    \Delta F \equiv \Delta F (\{ u_c^i \}, \{b_i^m \}, \{ p_m \}, \{ \mu_m \}, t_n)
    \label{eq:o_iProb}
\end{equation}
replacing $\Delta F$ in Eq.~\ref{eq:o_i}.

\begin{figure}[ht!]%
    \includegraphics[width=0.45\textwidth]{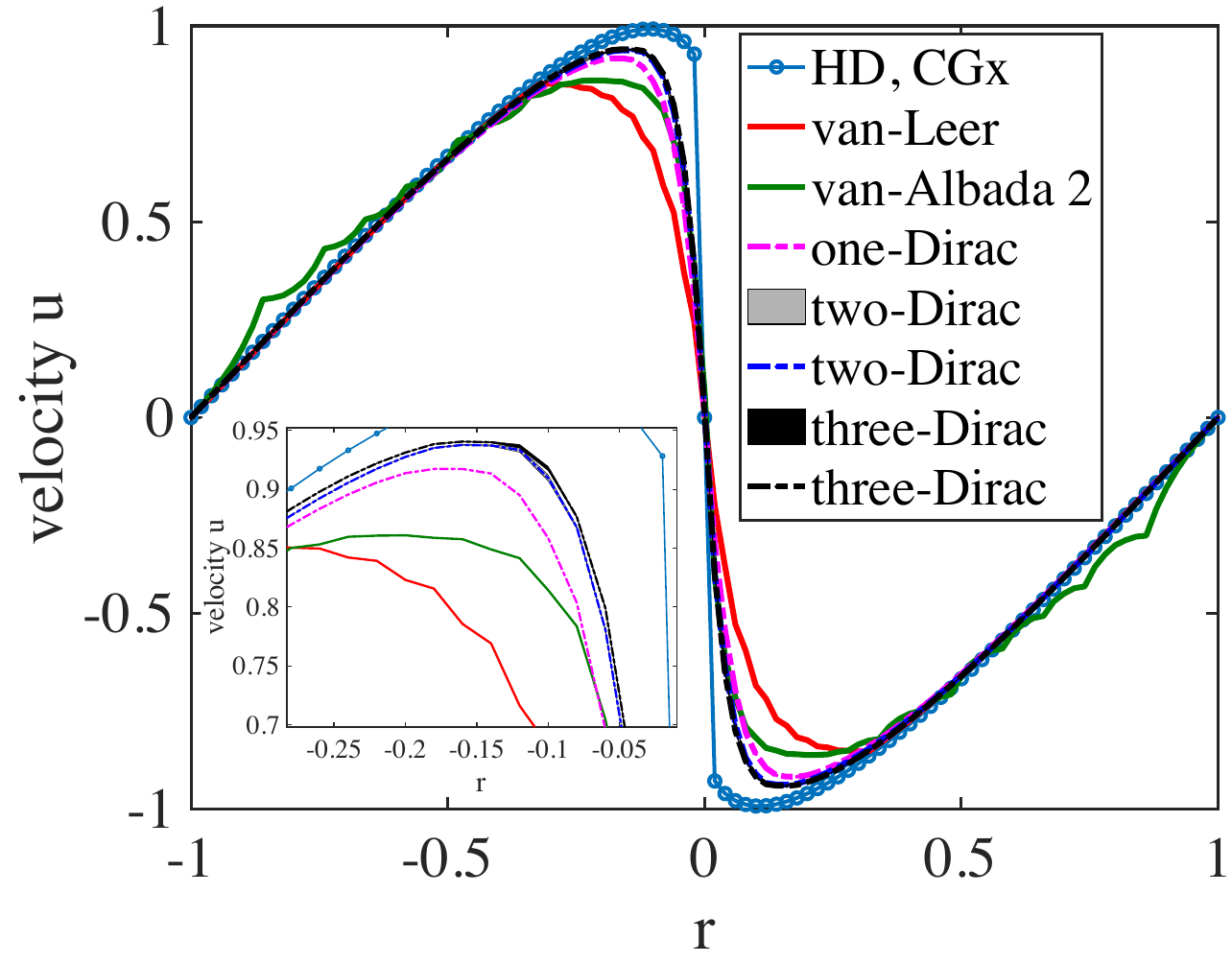}
    \caption[justification=left]{Shock reconstruction in the case of Burgers' equation learned with $N_D \in [1, 2, 3]$ Dirac delta functions as compared to van Leer and van Albada 2 cases.  Ground truth is plotted as blue connected circles (HD). The inset is a magnification of the solution at the upper left of the shock. Note that the probabilistic flux limiters in this figure are plotted as gray ($N_D=2$) and black ($N_D=3$) bands indicating mean plus and minus standard deviation. The probabilistic flux limiter with $N_D = 3$ used in this plot is given in Tables \ref{tab:threeDiracs_r} and \ref{tab:threeDiracs_b}. }
    \label{fig:onetwothreediracs_recon}
    \vspace{-0.15in}
\end{figure}

\COMMENT{
\begin{wrapfigure}{r}{0.2\textwidth}
  \vspace{-15pt}
\minipage{0.09\textwidth}
  \includegraphics[width=1\linewidth]{figure1}
  \includegraphics[width=1\linewidth]{figure2}
  \endminipage\hfill
\minipage{0.09\textwidth}
  \includegraphics[width=1\linewidth]{figure3}
  \includegraphics[width=1\linewidth]{figure4}
  \endminipage
  \hspace{2.0cm}
  \vspace{-5pt}
  \caption{Caption goes here}
  \label{fig:example}
  \vspace{-10pt}
\end{wrapfigure}
} 

\subsection{Dataset}
We trained our machine learning algorithm and attained good convergence using $50$ distinct Burgers' simulations evolved from $50$ random initial conditions that have in total $160$M training data points (note $N_c=6$ as  discussed above).  We held out $10$ simulations of $16$M data points for testing. Timestep, $dt$, and grid size, $dx$, associated with our high-definition simulations were, respectively, $1.0e^{-4}$ and $2.5e^{-3}$ and these two values were fixed for the ground-truth (high-resolution) simulations.  Our training dataset contained various viscosity values, $\mu \in [0.002, 0.03]$. Here, $x \in [-1, 1]$ and $t \in [0, 0.4]$.

\subsection{Probabilistic solutions}
%
\begin{table*}[h]
\begin{tabular}{|l|l|l|l|l|l|}
\hline
{\color[HTML]{333333} FLs}        & {\color[HTML]{333333} van Leer} & {\color[HTML]{333333} van Albada 2} & {\color[HTML]{333333} 1 Dirac}  & {\color[HTML]{333333} 2 Diracs} & {\color[HTML]{333333} 3 Diracs} \\ \hline
{\color[HTML]{333333} Mean Error} & {\color[HTML]{333333} 0.71 x $10^{-3}$ } & {\color[HTML]{333333} 0.0768 x $10^{-3}$ }     & {\color[HTML]{333333} 0.0762 x $10^{-3}$ } & {\color[HTML]{333333} 0.0758 x $10^{-3}$ } & {\color[HTML]{333333} 0.0753 x $10^{-3}$ } \\ \hline
\end{tabular}
\caption{Mean reconstruction errors ($\mathrm{e_{mean}}$) obtained on the test set (data not trained and random initial conditions) using machine learned probabilistic flux limiters ($N_D = 1,2,3$) as compared to van Leer and van Albada 2.  Learned flux limiters were optimized on the entire parameter range of $K=[2,3,\dots,38]$, $CG=[2,3,\dots,10]$, and all $\mu_i$'s are constrained to be in the interval $\mu_i=[0.002,0.03]$.}
\label{tab:CGoptimal_testset}
\end{table*}
%
\begin{table*}[h]
\begin{tabular}{|l|l|l|l|l|l|}
\hline
{\color[HTML]{333333} FLs}        & {\color[HTML]{333333} van Leer} & {\color[HTML]{333333} van Albada 2} & {\color[HTML]{333333} 1 Dirac}  & {\color[HTML]{333333} 2 Diracs} & {\color[HTML]{333333} 3 Diracs} \\ \hline
{\color[HTML]{333333} Mean Error} & {\color[HTML]{333333} 0.71 x $10^{-3}$ } & {\color[HTML]{333333} 0.0768 x $10^{-3}$ }     & {\color[HTML]{333333} 0.0764 x $10^{-3}$ } & {\color[HTML]{333333} 0.0761 x $10^{-3}$ } & {\color[HTML]{333333} 0.0754 x $10^{-3}$ } \\ \hline
\end{tabular}
\caption{Mean reconstruction errors ($\mathrm{e_{mean}}$) obtained on the test set (data not trained and random initial conditions) using machine learned probabilistic flux limiters ($N_D = 1,2,3$) as compared to van Leer and van Albada 2 for $CG=8$.  Learned flux limiters were optimized while constraining the coarse graining to $CG=8$ and the other parameters are in the ranges $K=[2,3,\dots,38]$, and $\mu_i=[0.002,0.03]$.}
\label{tab:CG8_testset}
\end{table*}
%
\begin{table*}[h]
\begin{tabular}{|l|l|l|l|l|l|}
\hline
{\color[HTML]{333333} FLs}        & {\color[HTML]{333333} van Leer} & {\color[HTML]{333333} van Albada 2} & {\color[HTML]{333333} 1 Dirac}  & {\color[HTML]{333333} 2 Diracs} & {\color[HTML]{333333} 3 Diracs} \\ \hline
{\color[HTML]{333333} Mean Error} & {\color[HTML]{333333} 0.926 x $10^{-3}$ } & {\color[HTML]{333333} 0.235 x $10^{-3}$ }     & {\color[HTML]{333333} 0.367 x $10^{-3}$ } & {\color[HTML]{333333} 0.194 x $10^{-3}$ } & {\color[HTML]{333333} 0.189 x $10^{-3}$ } \\ \hline
\end{tabular}
\caption{Mean reconstruction errors ($\mathrm{e_{mean}}$) obtained with sinusoidal initial condition using machine learned probabilistic flux limiters as compared to van Leer and van Albada 2.  Learned flux limiters were optimized on the entire parameter range of $K=[2,3,\dots,38]$, $CG=[2,3,\dots,10]$, and all $\mu_i$'s are constrained to be in the interval $\mu_i=[0.002,0.03]$.}
\label{tab:CGoptimal_sin}
\end{table*}
%
\begin{table*}[h]
\begin{tabular}{|l|l|l|l|l|l|}
\hline
{\color[HTML]{333333} FLs}        & {\color[HTML]{333333} van Leer} & {\color[HTML]{333333} van Albada 2} & {\color[HTML]{333333} 1 Dirac}  & {\color[HTML]{333333} 2 Diracs} & {\color[HTML]{333333} 3 Diracs} \\ \hline
{\color[HTML]{333333} Mean Error} & {\color[HTML]{333333} 0.926 x $10^{-3}$} & {\color[HTML]{333333} 0.235 x $10^{-3}$}     & {\color[HTML]{333333} 0.063 x $10^{-3}$} & {\color[HTML]{333333} 0.031 x $10^{-3}$} & {\color[HTML]{333333} 0.024 x $10^{-3}$} \\ \hline
\end{tabular}
\caption{Mean reconstruction errors ($\mathrm{e_{mean}}$) obtained with sinusoidal initial condition using machine learned probabilistic flux limiters as compared to van Leer and van Albada 2 for $CG=8$.  Learned flux limiters were optimized while constraining the coarse graining to $CG=8$ and the other parameters are in the ranges $K=[2,3,\dots,38]$, and $\mu_i=[0.002,0.03]$.}
\label{tab:CG8_sin}
\end{table*}
%

Fig.~\ref{fig:threediracsrun} captures the results of an optimization of a probabilistic flux limiter over a parameter space with $CG=[2,3, \dots, 10]$, $K=[2,3, \dots,38]$, and $\mu=[0.002, 0.03]$ with probabilities $p_i=[0,1]$ with the constraint $p = p_1 + p_2 + p_3 = 1$ and $N_D = 3$.  Mean and variance constraints defining the range of viscosities $\mu_m$ were tuned to be, respectively, $4.0e^{-3}$ and $1.0e^{-4}$.  Typically, stable solutions were attained (see Fig.~\ref{fig:threediracsrun}(a), (b), and (c)) after approximately $60$ iterations of the optimizer. 

The three blue curves in Fig.~\ref{fig:threediracsrun}(a) main plot and inset represent the $3$ corresponding pairs ($\mu_i$, $p_i$), where we use the same line style for each pair.  The dominant contributions to the optimized limiters came from ($\mu_1$, $p_1$) and ($\mu_2$, $p_2$) while the remaining pair ($\mu_3$, $p_3$) contributed at a lower probability to the solution.  

Explicit values for the endpoints of the linear segments and the slopes of the piecewise-linear flux limiting functions are presented in Tables ~\ref{tab:threeDiracs_r} and ~\ref{tab:threeDiracs_b} in  Appendix~\ref{sec:appendix_closedformsFL}. The optimization for the limiter in  Fig.~\ref{fig:threediracsrun} was performed over the entire computational parameter range, giving a limiter that worked best on average in this parameter space. An additional probabilistic flux limiter is depicted in Appendix \ref{sec:appendix_closedformsFL}, Fig.~\ref{fig:threediracs_CG8_optimization}. For the optimization in Fig.~\ref{fig:threediracs_CG8_optimization}, the coarse graining parameter was constrained to $CG=8$. Note the differences found toward the tail (large $r$) of the flux limiters in Fig.~\ref{fig:threediracsrun} and Fig.~\ref{fig:threediracs_CG8_optimization}, resulting from the tighter constraint on $CG$.

We studied  probabilistic flux limiters obtained for the cases $N_D \in [1, 2, 3]$ (i.e. sets of one, two or three flux limiting functions with associated selection probabilities) and compared these probabilistic flux limiters with van Leer and van Albada 2 (note that the $N_D = 1$ case was studied in \cite{NguyenMcKernsSornborger} and was shown to outperform $12$ flux limiters, including van Leer and van Albada 2, from the literature). In Tables (\ref{tab:CGoptimal_testset}, \ref{tab:CG8_testset}), we present mean squared errors (MSE) measuring the shock capturing prediction $\{o_i\}$ as compared to ground truth, high-resolution data, $\{g_i \}$,
\begin{equation}
  e_\mathrm{mean} = \frac{1}{2 N} \sum_{i=1}^{N} (o_i - g_i)^2 \; .
  \label{eq:error}
\end{equation}
In Tables (\ref{tab:CGoptimal_sin}, \ref{tab:CG8_sin}) we present MSE between a sinusoidal solution to Burgers' equation and the shock capturing prediction.


For $N_D > 1$, we obtain at least $2$ pronounced contributions to the total probability distribution.  For $N_D = 2$, our optimization yielded two Dirac delta functions with strong (i.e. high probability) contributions of $p_i\approx0.5$ each, while for $N_D = 3$ our optimization is shown in Fig.~\ref{fig:threediracsrun} with 2 flux limiters with strong contributions and one flux limiter with stable, but smaller probability, $p_3 \gtrapprox 0$.

\begin{table*}[]
\begin{tabular}{|c|c|c|l|c|l|}
\hline
$\mu$ & van Leer & van Albada 2 & 1 Dirac & 2 Diracs & 3 Diracs \\ \hline
0.002  & 2.33 x $10^{-3}$     & 0.60 x $10^{-3}$         & 0.62 x $10^{-3}$    & 0.46 x $10^{-3}$     & \textbf{0.42 x $10^{-3}$}    \\ \hline
0.00498  & 2.08 x $10^{-3}$     & 0.46 x $10^{-3}$         & 0.41 x $10^{-3}$   & 0.27 x $10^{-3}$     & \textbf{0.24 x $10^{-3}$}     \\ \hline
0.00625  & 1.89 x $10^{-3}$     & 0.44 x $10^{-3}$         & 0.37 x $10^{-3}$   & 0.20 x $10^{-3}$     & \textbf{0.17 x $10^{-3}$ }    \\ \hline
0.02  & 1.02 x $10^{-3}$     & 0.23 x $10^{-3}$         & \textbf{0.001 x $10^{-3}$}    & 0.03 x $10^{-3}$     & 0.046 x $10^{-3}$    \\ \hline
0.03  & 0.73 x $10^{-3}$     & 0.16 x $10^{-3}$         & \textbf{0.11 x $10^{-3}$}    & 0.25 x $10^{-3}$     & 0.29 x $10^{-3}$    \\ \hline
\end{tabular}
\caption{MSE (Eq. \ref{eq:error}) comparing high-resolution and flux limiter-based simulations averaged over $t = [0,0.4]$ and $x=[-1,1]$ for five different $\mu$ with sinusoidal initial conditions. Here, we compare machine learned probabilistic flux limiters ($N_D = 1,2,3$) to van Leer and van Albada 2 for a coarse-graining $CG = 8$. The learned probabilistic flux limiter was optimized over the ranges $K\in[2,3,\dots,38]$, and $\mu_i\in[0.002,0.03]$. The left column represents viscosities of a set of high-resolution (ground truth) simulations taken from the range of viscosities that the probabilistic flux limiter was trained on. Numbers in bold show superior performance. Note that all of the best performance results were for machine learned limiters. For small viscosities (upper three rows), probabilistic flux limiters with $N_D = 3$ were best able to capture the sharper shocks.}
\label{tab:errors_mus}
\end{table*}
\begin{table}[ht]
\begin{tabular}{cccc|cc|}
\hline
\multicolumn{2}{|c|}{$N_D=1$}                           & \multicolumn{2}{c|}{$N_D=2$}         & \multicolumn{2}{c|}{$N_D=3$}         \\ \hline
\multicolumn{1}{|c|}{$p$} & \multicolumn{1}{c|}{$\mu_1$}      & \multicolumn{1}{c|}{$p$}    & $\mu_{1-2}$      & \multicolumn{1}{c|}{$p$}    & $\mu_{1-3}$      \\ \hline
\multicolumn{1}{|c|}{1} & \multicolumn{1}{c|}{0.00625} & \multicolumn{1}{c|}{0.09} & 0.00562 & \multicolumn{1}{l|}{0.09} & 0.00511 \\ 
\hline
& \multicolumn{1}{l|}{}        & \multicolumn{1}{l|}{0.91} & 0.00492 & \multicolumn{1}{l|}{0.08} & 0.00439 \\ \cline{3-6} 
&                              &                           &         & \multicolumn{1}{l|}{0.83} & 0.00502 \\ \cline{5-6} 
\end{tabular}
\caption{Optimal selection probabilities, $p_i$, and associated viscosities, $\mu_i$, of probabilistic flux limiter functions used in Tab.~\ref{tab:errors_mus}}
\label{tab:123D_probandmu}
\end{table}

In Table~\ref{tab:errors_mus}, we evaluated MSE obtained from five characteristic  viscosities, $\mu$ (left column), on test cases with sinusoidal initial conditions, using machine-learned probabilistic flux limiters (for $N_D=1,2,3$).  These errors are compared to traditional van Leer and van Albada 2 methods, with a coarse graining set to $CG=8$. The optimization of learned flux limiters was thus constrained by $CG=8$, with other parameters varying within $K=[2,3,…,38]$ and $\mu_i =[0.002,0.03]$.  We used $CG=8$, corresponding to $dt=8 \times 1.0e^{-4}$ and $dx=8 \times 2.5e^{-3}$.  The lower bound case of $\mu=0.002$ is depicted in the top row. The viscosity ($\mu=0.00498$) was average of the three associated viscosities obtained with the $N_D=3$ flux limiter. This was also the case with $N_D=2$ flux limiters. The next viscosity ($\mu=0.00625$) was derived from $N_D=1$ flux limiter, and the final two $\mu$ values were selected from the lower and upper bounds of the viscosities used over which we trained the probabilistic flux limiter.  Details of the probabilistic flux limiters' optimized characteristic parameters are shown in Tab.~\ref{tab:123D_probandmu}. 


In Fig.~\ref{fig:onetwothreediracs_recon}, we plot solutions to the analytically solvable sine wave problem obtained using probabilistic flux limiters for $N_D \in [1, 2, 3]$ in comparison to ground truth (blue connected circles), van Leer, and van Albada 2 for $\mu=0.002$. Note that this plot captures a sufficiently large time such that the shock has evolved to be sharp. All machine learned probabilistic flux limiters outperformed van Leer and van Albada 2 limiters, with the $N_D = 3$ case (black dashed line) performing with the highest accuracy. This result was consistent across all considered values of $\mu$, spanning the entire range from lower bound to upper bound, as detailed in Table~\ref{tab:errors_mus}. 

It is important to note that when $N_D=1$, the learned limiter performed very well for the larger viscosity band, but had a marginally lower performance compared to van Albada 2 across the entire time interval $[0,0.4]$ at small viscosities. Specifically when $\mu=0.002$, van Albada 2 provided a slightly better solution than $N_D=1$ at early times before shocks occurred. For $N_D = 1, 2, 3$ the same optimized solutions were employed across all study cases. In contrast, distinct solutions utilizing van Leer and van Albada 2, along with corresponding ground truths, were derived for each $\mu$. Thus, we expect that there should be one viscosity for which the $N_D = 1$ limiter should match very well the ground truth data. On the other hand, the $N_D > 1$ limiters exhibit smaller overall errors over the whole range of viscosities and greater generalizibility potential when the viscosity is scaled by the ground truth viscosity (see below). 


\subsection{Robustness and Generalizability}
\label{seq:gen}
In Table~\ref{tab:errors_mus}, we studied a fixed probabilistic flux limiter learned across various viscosities $\mu$. In Appendix~\ref{sec:appendix_closedformsFL}, Table~\ref{tab:errors_mus_various}, we show that, by scaling the $\mu_i$ values associated with a probabilistic flux limiter with $\mu$ value of the high-resolution simulation, we can further extend the domain of application of the limiter. This demonstrates the benefit of viscosity scaling outside the optimal range expected from our machine learning procedure.

The $N_D=3$ case consistently showed the best performance: mean profile indicating a sharper shock than the deterministic limiters with smaller standard deviation.   
Even though we plotted standard deviation bands taken from $100$ test runs, the widths of the bands were barely distinguishable from the unaveraged lines plotted for the non-probabilistic flux limiters. 

The $N_D=3$ case, in both Tabs.~\ref{tab:errors_mus} and~\ref{tab:errors_mus_various}, slightly outperformed the $N_D=2$ case, while the traditional flux limiters of van Leer or van Albada 2, i) had lower accuracy than that produced with machine learned probabilistic flux limiters, and ii) showed ``kinks" in the shock reconstructions as compared to the smooth shock reconstructions obtained by our probabilistic limiters. 
Kinks occurred when $\mu$ was sufficiently small. As a control run, we examined van Albada 2 solutions with larger $\mu$ values (e.g. $\mu=0.1$, larger viscosity) and these ``oscillations" disappeared.  

The improvement in performance going from $N_D=2$ to $N_D=3$ was significantly less than the improvement going from $N_D=1$ to $N_D=2$. This (and consistently obtaining $p_3 \gtrapprox 0$) suggests that the $N_D=3$ Dirac delta function case well approximates a distribution that is sufficient to produce an optimal probabilistic limiter for the system under consideration.

\section{Discussion}
In this paper, we presented a conceptually new type of flux limiter that we refer to as a {\it probabilistic flux limiter} since its use consists of drawing randomly from a set of flux limiting functions, $\phi_m \coloneqq \phi_m(CG, K, \mu)$, optimized on the parameter hypercube defined by $(CG, K, \mu, p)$. 

We quantified the effectiveness of machine learned probabilistic flux limiters for integrating a coarse-grained, one-dimensional Burgers' equation in time.  Probabilistic flux limiters were trained on coarse-grained data taken from a high resolution dataset with random initial conditions.  With the learned probabilistic flux limiter, we then integrated on unseen cases of both random and sinusoidal initial conditions.  

Our results consistently showed that the learned probabilistic flux limiters can more accurately capture the overall coarse-grained evolution of the flow, and, in particular, shock formation relative to conventional flux limiters, e.g. van Leer and van Albada 2.  

Note that in Fig.~\ref{fig:threediracsrun}(b), the optimal coarse-graining over all possible coarse-grainings was $2$ for this case. This should be considered as distinct from fixing $CG = 2$ and optimizing over other parameters. Further, the optimal $K$, in this case, was at the upper bound of the parameter space. By extending the parameter space, we could potentially have found better performance, but most of the segments in the limiter were found in the range $r \approx 1$, and it was likely that they would only marginally improve the limiter that was found.

The improvement that one finds as $N_D$ is incremented decreases. This demonstrates that although there was significant improvement, there were also diminishing returns as $N_D$ was increased from $2$ to $3$.

We learned probabilistic flux limiters for $N_D$ up to $3$, which seemed to provide a sufficient number of Dirac delta functions to approximate the optimal probabilistic flux limiter for the system studied in this manuscript. 
The results obtained in this paper suggest potential applications of probabilistic flux limiters for better shock capture in more complex flow simulations.

\section{Acknowledgements}
Research presented in this article was supported by the NNSA Advanced Simulation and Computing Beyond Moore's Law Program at Los Alamos National Laboratory, and by the Uncertainty Quantification Foundation under the Statistical Learning program. Los Alamos National Laboratory is operated by Triad National Security, LLC, for the National Nuclear Security Administration of the U.S. Department of Energy (Contract No. 89233218CNA000001). The Uncertainty Quantification Foundation is a nonprofit dedicated to the advancement of predictive science through research, education, and the development and dissemination of advanced technologies. This document's LANL designation is LA-UR-24-22187. 

\bibliographystyle{unsrt}
\bibliography{Biblio.bib}

\appendix
\setcounter{table}{0}
\renewcommand{\thetable}{A\arabic{table}}
\section{Parameters of optimized flux limiters} \label{sec:appendix_closedformsFL}
Closed forms of the flux limiting functions (iteration $65$) for $N_D = 3$ are found in Tables ~\ref{tab:threeDiracs_r} and ~\ref{tab:threeDiracs_b}. 

\begin{table*}[t]
{\scriptsize
\centering
\begin{tabular}{|c|c|c|c|c|c|c|c|c|c|c|c|c|c|c|c|c|c|c|c|c|c|c|} 
\hline
    & $r_1$   & $r_2$    & $r_3$   & $r_4$   & $r_5$   & $r_6$   & $r_7$   & $r_8$   & $r_9$   & $r_{10}$  & $r_{11}$  & $r_{12}$  & $r_{13}$  & $r_{14}$   & $r_{15}$   & $r_{16}$   & $r_{17}$   & $r_{18}$  & $r_{19}$  & $r_{20}$  & $r_{21}$  & $r_{22}$   \\ 
\cline{2-23}
    & $r_{23}$   & $r_{24}$    & $r_{25}$   & $r_{26}$   & $r_{27}$   & $r_{28}$   & $r_{29}$   & $r_{30}$   & $r_{31}$   & $r_{32}$   & $r_{33}$   & $r_{34}$   & $r_{35}$   & $r_{36}$   & $r_{37}$    & $r_{38}$    &  $r_{39}$    &      &      &      &      &       \\
\hline
65 & 0.0  & 0.23  & 0.38 & 0.49 & 0.57 & 0.64 & 0.69 & 0.73 & 0.77 & 0.80 & 0.83 & 0.86 & 0.88 & 0.90  & 0.92  & 0.94  & 0.96  & 0.97 & 0.99 & 1.00 & 1.01 & 1.05  \\
\cline{2-23}
    & 1.07 & 1.09  & 1.11 & 1.13 & 1.16 & 1.13 & 1.19 & 1.22 & 1.27 & 1.33 & 1.41 & 1.52 & 1.69 & 1.96  & 2.51 &  4.15 & 10.00      &      &      &      &      &      \\
\hline
65 & 0.0  & 0.24  & 0.40 & 0.51 & 0.59 & 0.65 & 0.70 & 0.74 & 0.78 & 0.81 & 0.84 & 0.86 & 0.89 & 0.91  & 0.93  & 0.94  & 0.96  & 0.98 & 0.99 & 1.00 & 1.02 & 1.04  \\
\cline{2-23}
    & 1.05 & 1.07  & 1.09 & 1.11 & 1.13 & 1.15 & 1.18 & 1.22 & 1.26 & 1.32 & 1.40 & 1.50 & 1.66 & 1.92  & 2.45 & 4.03 &  10.00     &      &      &      &      &      \\
\hline
65 & 0.0  & 0.24  & 0.40 & 0.51 & 0.59 & 0.65 & 0.70 & 0.74 & 0.78 & 0.81 & 0.83 & 0.86 & 0.89 & 0.91  & 0.93  & 0.94  & 0.96  & 0.98 & 0.99 & 1.00 & 1.02 & 1.03  \\
\cline{2-23}
    & 1.05 & 1.06  & 1.08 & 1.10 & 1.12 & 1.15 & 1.18 & 1.21 & 1.26 & 1.31 & 1.39 & 1.49 & 1.65 & 1.91  & 2.42 & 3.98  &  10.00     &      &      &      &      &      \\
\hline
\end{tabular}
}
\caption{ Coordinates of $3$ flux limiters in the case $N_D  = 3$.   Results obtained over the entire search range of $CG=[2,3, \dots, 10], K=[2,3, \dots, 38],$ and $\mu_i$'s are constrained to be in the interval $\mu_i=[0.002,0.03]$. } 
\label{tab:threeDiracs_r}
\end{table*}
\begin{table*}[th]
\scriptsize{
\centering
\begin{tabular}{|c|c|c|c|c|c|c|c|c|c|c|c|c|c|c|c|c|c|c|c|c|c|} 
\hline
    & $b_1$    & $b_2$    & $b_3$    & $b_4$    & $b_5$    & $b_6$    & $b_7$    & $b_8$    & $b_9$    & $b_{10}$   & $b_{11}$   & $b_{12}$   & $b_{13}$   & $b_{14}$   & $b_{15}$   & $b_{16}$   & $b_{17}$   & $b_{18}$   & $b_{19}$   & $b_{20}$   & $b_{21}$  \\ 
\cline{2-22}
    & $b_{22}$     & $b_{23}$    & $b_{24}$    & $b_{25}$    & $b_{26}$    & $b_{27}$    & $b_{28}$    & $b_{29}$    & $b_{30}$    & $b_{31}$   & $b_{32}$   & $b_{33}$   & $b_{34}$   & $b_{35}$   & $b_{36}$   &  $b_{37}$  &  $b_{38}$  &       &       &       &    \\
\hline
65 & 5.14  & -3.01 & 1.11  & -0.46 & 0.27  & 0.28 & 0.16  & 0.15 & -0.06  & 0.25 & 0.26  & -0.27 & 0.39  & -0.19 & 0.15  & 0.42 & 0.10  & -0.22 & 0.46 & -0.11  & 0.15        \\
\cline{2-22}
     & 0.07   & 0.24 & -0.46  & 0.58  & -0.25  & 0.41 & 0.01  & 0.02 & 0.19  & 0.02 & 0.18 & 0.12  & 0.01  & 0.27 &  0.24 & 0.34 &    -0.14   &       &       &       &      \\
\hline
65 & 4.93  & -2.88 & 1.04  & -0.45 & 0.55  & -0.05 & 0.20  & 0.07 & 0.38  & -0.38 & 0.31   & 0.61 & -0.47  & 0.05 & 0.45  & 0.20 & -0.29  & 0.09 & -0.05 & 0.10  & 0.60       \\
\cline{2-22}
     & -0.60   & 0.56 & -0.04  & -0.28 & 0.80  & -0.61 & 0.43  & 0.26 & -0.03  & 0.09 & 0.30 & 0.03  & 0.10  & 0.22 &  0.04 & 0.32 & -0.15      &       &       &       &      \\
\hline
65 & 4.86  & -2.90 & 1.24  & -0.55 & 0.47  & -0.08 & 0.13  & 0.17 & -0.08  & 0.42 & 0.39  & 0.19 & -0.46  & 0.53 & 0.05  & 0.10 & -0.07  & 0.15 & -0.01 & 0.31  & 0.07        \\
\cline{2-22}
     & 0.02   & 0.05 & 0.33  & -0.20  & 0.16  & -0.32 & 0.36  & 0.01 & 0.10  & 0.19 & 0.03 & 0.20  & 0.08  & 0.30 &  -0.05 & 0.32 &    -0.14   &       &       &       &      \\
\hline
\end{tabular}
}
\caption{ Slopes of flux limiters in the case $N_D = 3$.  Results obtained over the entire search range of $CG = [2,3, \dots, 10], K = [2,3, \dots, 38]$, and $\mu_i = [0.002, 0.03]$. $p_i$'s are in the interval $p_i = [0, 1]$ obeying unit sum for probability. }
\label{tab:threeDiracs_b}
\end{table*}
%
%

\setcounter{figure}{0}    
\renewcommand{\thefigure}{A\arabic{figure}}  
We show in Fig.~\ref{fig:threediracs_CG8_optimization} the flux limiting functions and probabilities (lower inset) at the termination of the optimization procedure for $N_D=3$ with the constraint $CG=8$. The yellow curve in the upper inset of Fig.~\ref{fig:threediracs_CG8_optimization} represents the change in bin number $K$ along with corresponding viscosity $\mu_1, \mu_2, \mu_3$ and probability $p_1, p_2, p_3$, respectively, in thick, medium, and thin solid lines (both insets).
%
\begin{figure}[!h]%
    \includegraphics[width=0.4\textwidth]{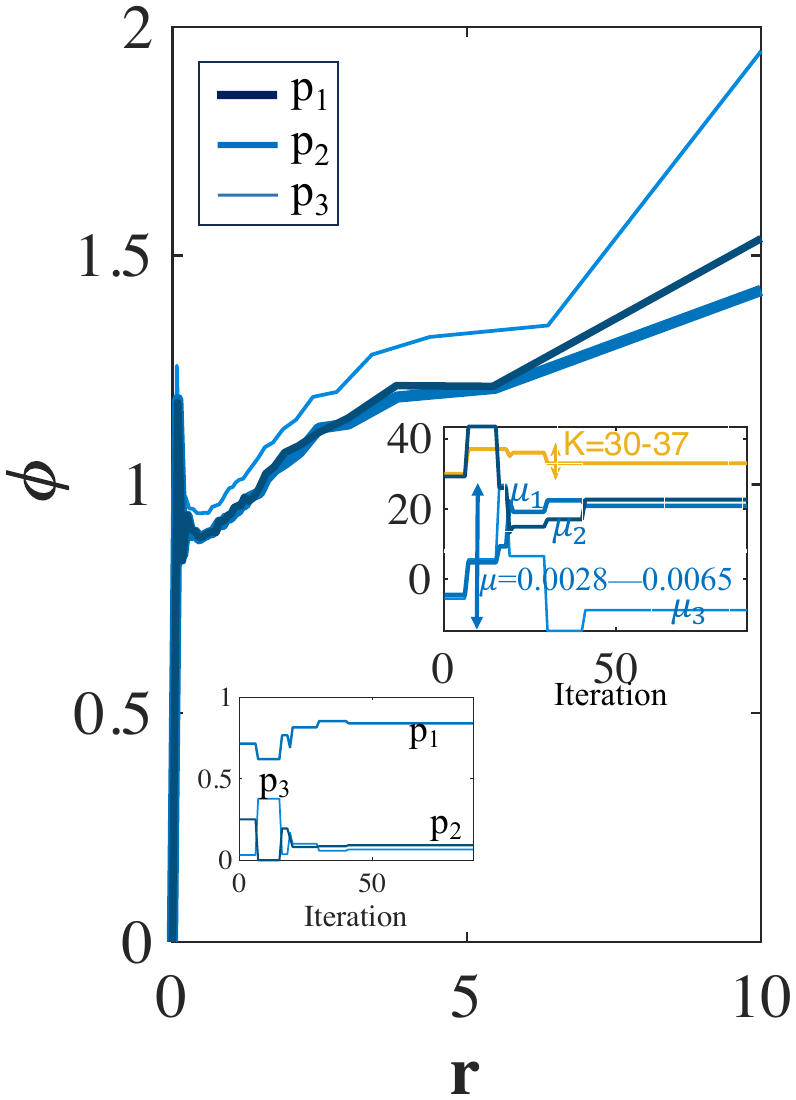}
    \caption[justification=left]{Probabilistic flux limiter with coarse graining constraint. A probabilistic flux limiter obtained with $N_D = 3$ and constraint $CG=8$.  Top inset is parameter convergence with respect to optimization iteration and bottom inset is probability $p_1, p_2, p_3$ corresponding to the weights of the $3$ Dirac delta functions as a function of optimization iteration.}
    \label{fig:threediracs_CG8_optimization}
\end{figure}
%

The errors compared against several high-resolution simulations with different viscosities are shown in Table ~\ref{tab:errors_mus_various}. In this case, the training was performed with the same viscosity as the high-resolution case. For this application, increasing the number of Diracs in the description of the ML limiter always leads to lower errors.

\begin{table*}[]
\begin{tabular}{|c|c|c|c|c|c|}
\hline
$\mu$ & van Leer & van Albada 2 & 1 Dirac & 2 Diracs & 3 Diracs \\ \hline
0.002    & 2.33 x $10^{-3}$     & 0.60 x $10^{-3}$            & 0.49 x $10^{-3}$     & 0.32 x $10^{-3}$        & \textbf{0.29 x $10^{-3}$}    \\ \hline
0.00498  & 2.08 x $10^{-3}$     & 0.46 x $10^{-3}$           & 0.41 x $10^{-3}$     & 0.27 x $10^{-3}$      & \textbf{0.24 x $10^{-3}$}     \\ \hline
0.00625  & 1.89 x $10^{-3}$     & 0.44 x $10^{-3}$       & 0.037 x $10^{-3}$   & 0.024 x $10^{-3}$     & \textbf{0.021 x $10^{-3}$}    \\ \hline
0.02     & 1.02 x $10^{-3}$     & 0.23 x $10^{-3}$            & 0.19 x $10^{-3}$      & 0.11 x $10^{-3}$       & \textbf{0.08 x $10^{-3}$}    \\ \hline
0.03     & 0.73 x $10^{-3}$     & 0.16 x $10^{-3}$         & 0.14 x $10^{-3}$    & 0.08 x $10^{-3}$     & \textbf{0.06 x $10^{-3}$}    \\ \hline
\end{tabular}
\caption{ MSE obtained for five different viscosities $\mu$, similar values as shown in Tab.~\ref{tab:errors_mus}, on the test case of a sinusoidal initial condition using machine learned probabilistic flux limiters ($N_D = 1,2,3$) as compared to van Leer and van Albada 2 for $CG=8$. Optimal flux limiters were learned while constraining the coarse graining to $CG=8$ and the other parameters are in the ranges $K=[2,3,\dots,38]$, and $\mu_i=[0.002,0.03]$. Unlike the results shown in Tab.~\ref{tab:errors_mus} for $N_D=1, 2, 3$ Dirac delta functions (where the learned flux limiters were trained at different $\mu$ than they were applied),
here, these MSE were obtained with the limiter trained for the $\mu$ in the far left column. Bold numbers indicate superior performance. }
\label{tab:errors_mus_various}
\end{table*}

\end{document}